\documentclass[aps,prl,twocolumn,superscriptaddress,preprintnumbers,footinbib]{revtex4-1}
\usepackage{amssymb}
\usepackage{amsmath}
\usepackage{epsfig}
\usepackage{hyperref}
\usepackage{breakurl}
\usepackage{cleveref}
\usepackage{color}
\usepackage{url}

\usepackage{bm}
\usepackage[bbgreekl]{mathbbol}
\DeclareMathSymbol\bbDelta  \mathord{bbold}{"01}

\usepackage{graphicx}
\usepackage[export]{adjustbox}

\makeatletter
\def\simgt{\mathrel{\lower2.5pt\vbox{\lineskip=0pt\baselineskip=0pt
           \hbox{$>$}\hbox{$\sim$}}}}
\def\simlt{\mathrel{\lower2.5pt\vbox{\lineskip=0pt\baselineskip=0pt
           \hbox{$<$}\hbox{$\sim$}}}}
\makeatother

\Crefname{equation}{Eq.}{Eqs.}

\newcommand{\be}{\begin{equation}}
\newcommand{\ee}{\end{equation}}
\newcommand{\bea}{\begin{eqnarray}}
\newcommand{\eea}{\end{eqnarray}}
\newcommand{\eq}[2]{\be\begin{aligned}#1 \label{#2}\end{aligned}\ee}

\newcommand{\Reff}[1]{Ref.~\cite{#1}}
\newcommand{\Fig}[1]{Fig.~\ref{#1}}
\newcommand{\Eq}[1]{Eq.~\eqref{#1}}

\newcommand{\tpartial}{\tilde\partial}

\newcommand{\ov}{\overline}

\renewcommand{\b}[1]{\bm{#1}}

\newcommand{\fddu}[3]{f_{#1#2}^{\phantom{#1 #2} #3}}
\newcommand{\fbarddu}[3]{f_{\ov #1 \ov #2}^{\phantom{\ov#1 \ov #2} \ov #3}}

\def\Section#1{\smallskip \noindent {\bf #1}}

\begin{document}

\title{Non-perturbative Double Copy in Flatland}

\author{Clifford Cheung}
\affiliation{Walter Burke Institute for Theoretical Physics, California Institute of Technology, Pasadena, California 91125}
\author{James Mangan}
\affiliation{Department of Physics and Astronomy, Northwestern University, Evanston, Illinois 60208}
\author{Julio Parra-Martinez}
\affiliation{Walter Burke Institute for Theoretical Physics, California Institute of Technology, Pasadena, California 91125}
\author{Nabha Shah}
\affiliation{Walter Burke Institute for Theoretical Physics, California Institute of Technology, Pasadena, California 91125}

\begin{abstract}

%We derive a non-perturbative, Lagrangian-level formulation of the double copy in two dimensions based on the fact that the $N\rightarrow \infty$ limit of the $U(N)$ algebra is the diffeomorphism algebra.  This furnishes a field theoretic definition of the double copy for a very broad class of scalar theories, including masses and higher-dimension operators.  An immediate corollary is the familiar amplitudes-level double copy at all orders in perturbation theory.  Remarkably, the double copy maps integrable theories onto one another, thus implying an isomorphism between their Lax connections, Wilson lines, and infinite towers of conserved currents.  We also present a stringy version of this framework in which finite $N$ defines the string tension.  Last but not least, we implement and verify the double copy at the level of non-perturbative classical solutions, both analytically and numerically.

We derive a non-perturbative, Lagrangian-level formulation of the double copy in two spacetime dimensions.  Our results elucidate the field theoretic underpinnings of the double copy in a broad class of scalar theories which can include masses and higher-dimension operators.  An immediate corollary is the amplitudes-level double copy at all orders in perturbation theory. Applied to certain integrable models, the double copy defines an isomorphism between Lax connections, Wilson lines, and infinite towers of conserved currents.  We also implement the double copy at the level of non-perturbative classical solutions, both analytically and numerically, and present a generalization of the double copy map that includes a fixed tower of higher-dimension corrections given by the Moyal algebra.

\end{abstract}

\preprint{CALT-TH 2022-015}

\maketitle

\Section{Introduction.}  Recent breakthroughs in scattering theory have unveiled an extraordinary hidden structure lying dormant within the fundamental laws of nature.
The so-called ``double copy'' \cite{BCJ1, BCJ2, BCJReview, KLT} is a mathematical formula, only proven at tree-level, that very simply relates {\it perturbative scattering amplitudes} of gravitons in Einstein's general relativity (GR) to those of gluons in Yang--Mills theory (YM). 

At a purely practical level, the double copy is an immensely efficient tool for recycling past results in gauge theory to derive new ones for gravity.  This approach has made feasible many previously intractable calculations, for example those relevant to the finiteness of supergravity theories \cite{sugra1, sugra2, sugra3, sugra4, sugra5, sugra6, sugra7, sugra8} and more recently, post-Minkowskian computations for black hole binary dynamics \cite{PM1, PM2, PM3, PM4} which are directly relevant to the LIGO experimental program \cite{Buonnano1, Buonnano2} and are, within the last three years, competing with the state of the art.

At the conceptual level, the double copy remains deeply mysterious.  Its structure transcends gauge theory and gravity and applies to a broad web of theories \cite{BCJReview}.  For example, the exact same double copy {\it also relates} all tree-level amplitudes of pions in the chiral limit to those of certain hypothetical scalars known as Galileons, which have been studied independently as viable theories of cosmology and modified gravity.

In broad strokes, the double copy maps gauge theory to gravity by first expressing every gauge theory amplitude as a sum over cubic graphs,
\eq{
A_n = \sum \limits_{{\rm cubic}} \frac{c_i n_i}{d_i},
}{}
where the $c_i$ are color factors (structure constants), the $n_i$ are kinematic numerators, and the $d_i$ are propagators \cite{BCJ1, BCJ2,BCJReview, ElvangHuang, JJTasi}.
{\it Color-kinematics duality} states that there exists a rearrangement of terms such that the kinematic numerators obey the same Jacobi identities as the color factors.
Gravity -- as the {\it square} or {\it double copy} of gauge theory -- is simply obtained by replacing each color factor with the associated kinematic numerator, $c_i \to n_i$.
%More details can be found in the original papers \cite{BCJ1, BCJ2} as well as the reviews \cite{BCJReview, ElvangHuang, JJTasi} and the references therein.

The double copy is an established fact about flat space, perturbative scattering amplitudes but its generality is far from understood.  To what extent does it apply off-shell \cite{SDYM1, SDYM2, XYZPaper, CCK, HenrikCS, Moynihan, NANS}?  To curved geometries \cite{AdS_CK, RaduAdS, AllicAds, Diwakar:2021juk, Farrow:2018yni, Lipstein:2019mpu, Zhou:2021gnu, Alday:2021odx, Albayrak:2020fyp,  Armstrong:2020woi, Alday:2022lkk}?  Non-perturbatively?  Finding answers to these questions could provide a non-perturbative, background independent mapping between gravity and far simpler quantum field theories.

%Ordinary double copy only proven at tree level, full perturbation theory, bearing on all double-copy constructions.

In this paper, we present a non-perturbative double copy in two spacetime dimensions.
This is the first off-shell Lagrangian level formulation of the double copy for interacting theories. \footnote{Remarkably, three-dimensional Chern-Simons theory -- which is {\it topological} -- automatically manifests off-shell, Lagrangian-level color-kinematics duality \cite{HenrikCS}.}
%\footnote{An off-shell double copy is known for a topological theory \cite{HenrikCS}.}
%exists for topological theories
Extending the proof of the double copy from tree level to all loop orders has implications for the understanding of all double copy constructions.
Our approach is inspired by a remarkable isomorphism between the algebras of unitary transformations and diffeomorphisms \cite{Hoppe},
\eq{
%\begin{array}{c}
%\textrm{color} \\
%\textrm{algebra} \\
%\end{array} 
%\leftrightarrow
%\begin{array}{c}
%\textrm{kinematic} \\
%\textrm{algebra} \\
%\end{array} 
\lim_{N\rightarrow\infty} U(N)
\sim
%\begin{array}{c}
%\textrm{Poisson}% \\
%%\textrm{algebra} \\
%\end{array} = 
\begin{array}{c}
\textrm{Diff}_{S^1\times S^1},
%\textrm{algebra} \\
\end{array} 
}{algebra_equivalence}
and applies to an enormous class of scalar theories, including masses and higher-dimension operators.

We apply this construction successively to map bi-adjoint scalar (BAS) theory to Zakharov-Mikhailov (ZM) theory \cite{ZM} to the special Galileon (SG) \cite{EFTFromSoft, EFTRecursion, Kurt_sGal}, thus deriving the corresponding and more familiar amplitudes-level double copy at all orders in perturbation theory. 
\footnote{In relation to the prototypical double copy described above, ZM plays the role of the gauge theory and SG is analogous to gravity.  BAS has the same role in both setups.}
Since ZM theory is classically integrable, it furnishes a Lax connection whose Wilson lines define an infinite tower of conserved currents, all of which are shown to double copy into corresponding objects in the SG.
%A stringy version of the double copy is also presented, where $N$ is finite and identified with the string tension parameterizing an infinite tower of higher-dimension operators.
An extension of the double copy based on the Moyal algebra is presented where $N$, the rank of the gauge group, parameterizes an infinite tower of higher-dimension operators.
\footnote{The Moyal algebra has appeared before in maps from non-commutative gauge theory to ordinary gauge theory \cite{Seiberg:1999vs, Floratos:2005ij}.  Gravity is notably missing from this picture so an immediate connection to the double copy is opaque but potentially promising nonetheless.  We thank a referee for bringing this to our attention.}
Note that at the classical level, ZM theory is very closely related to self-dual Yang-Mills (SDYM) theory \cite{SDYM1, SDYM2}, which exhibits identical integrable and Moyal structures \cite{Chacon:2020fmr}.

Implementing the double copy on non-perturbative, large-field configurations, we show analytically that every classical solution of the SG theory is isomorphic to corresponding dual solutions in ZM and BAS theory.  As a highly non-trivial check, we compute an explicit, large-field, numerical solution for soliton scattering in the SG theory, map it to a corresponding configuration in ZM theory for $U(N)$ at large $N$, and verify that it satisfies the ZM equations of motion to high precision.

\Section{Color Algebra.} 
A field  in the adjoint of $U(N)$ is a  Hermitian matrix, $\b V = V^a \b T_a$, where $[ \b T_b ]_a^{\phantom{a} c} = i \fddu{a}{b}{c}$ and $
{} [\b T_a, \b T_b] = i\fddu{a}{b}{c} \b T_c$.
For odd $N$ there exists a basis of generators $\b T_p$ labeled by a two-vector, $p \in \mathbb{Z}_N\times \mathbb{Z}_N$ \cite{Hoppe}.   In this basis,
%\eq{
$\b V = V^p \b T_p$
%,
%}{V_color}
where $V^{p*} = V^{-p}$, and
\eq{
{} [\b T_{p_i} , \b T_{p_j} ] &= i \fddu{p_i}{p_j}{p_k}\b T_{p_k},
}{Tp}
with the corresponding color structure constant,\footnote{In our conventions, the Minkowski metric and Levi-Civita tensor obey $\eta_{00}=\epsilon_{01} =1$ so that $\epsilon_{\mu\nu} \epsilon_{\rho\sigma} = -(\eta_{\mu\rho}\eta_{\nu\sigma}-\eta_{\mu\sigma}\eta_{\nu\rho})$.  
Furthermore, we define the dual derivative $\tpartial^\mu = \epsilon^{\mu\nu} \partial_\nu$
and the antisymmetric product $\langle ij\rangle = \epsilon^{\mu\nu} p_{i\mu} p_{j\nu}$.}
\eq{
\fddu{p_i}{p_j}{p_k} &= -\tfrac{N}{2\pi} \sin\left( \tfrac{2\pi}{N} \langle ij\rangle
\right) \delta_{p_i+p_j,p_k } \overset{N\rightarrow \infty}{=}-\langle ij\rangle \delta_{p_i+p_j,p_k }.
}{fppp_color}
Hence, the $N\rightarrow \infty$ limit
literally defines the algebra of volume-preserving diffeomorphisms on the torus \cite{Hoppe}, or equivalently, the Poisson algebra.  The toroidal geometry arises because the generator labels $p$ are defined mod $N$.

\Section{Kinematic Algebra.} \Eq{algebra_equivalence} implies that fields in the adjoint of $U(N)$ at large $N$ are isomorphic to field-dependent diffeomorphisms,
\eq{
\b V =   \epsilon^{\mu\nu} \partial_\mu V \partial_\nu  = \partial_\mu V \tpartial^\mu = -\tpartial^\mu V \partial_\mu,
}{}
which are volume-preserving because $\partial_\mu \tpartial^\mu V=0$.
This algebra is closed since the commutator of diffeomorphisms yields another diffeomorphism via
\eq{
\b Z = [\b V, \b W]  =[ \partial_\mu V \tpartial^\mu, \partial_\nu W \tpartial^\nu] = \partial_\mu Z \tpartial^\mu,
}{}
where $Z = \partial_\mu V \tpartial^\mu W$.   Motivated by these structures, we propose a color-kinematic duality replacement,
\eq{
V^a &\quad \rightarrow \quad V  \\
 \fddu{a}{b}{c} V^a W^b &\quad \rightarrow \quad   \partial_\mu V \tpartial^\mu W  \\
 g_{ab} V^a W^b &\quad \rightarrow \quad \int VW .
}{replacement_rules}
The first line simply maps any color-adjoint field to a corresponding singlet field.  The second line maps the color structure constant to a kinematic structure constant whose momentum space representation is
\eq{
\fddu{p_i}{p_j}{p_k} = -\langle ij\rangle \delta^2(p_i +p_j-p_k).
}{fppp_ZM}
This is literally the continuum limit of \Eq{fppp_color}, in accordance with the algebra isomorphism in \Eq{algebra_equivalence}.
The third line is obtained from the Killing form of $U(N)$ at large $N$, which effectively defines a Killing form for the diffeomorphism algebra \cite{Hoppe}.

%Under the replacements in \Eq{replacement_rules}, a product of color generators, $\b T_{a_1} \b %T_{a_2}\cdots 
%\b T_{a_n} =\fddu{b_1}{a_1}{b_2} \fddu{b_2}{a_2}{b_3}\cdots 
%\fddu{b_n}{a_n}{b_{n+1}}$, maps to a momentum space form factor,
%\fddu{\ell,}{p_1}{\ell + p_1} \fddu{\ell+p_1,}{p_2}{\ell + p_1+p_2} \cdots 
% \fddu{\ell+\cdots + p_{n-1},}{p_n}{\ell +\cdots + p_n}
%}{}
%\end{widetext}
%where similar to \cite{Cheung:2022vnd}, repeated indices denote integration over internal momenta which localize momentum-conserving delta functions, yielding objects with the structure of a tree-level kinematic numerators.

\Section{Lagrangian Double Copy.} The color-kinematic replacement rules in \Eq{replacement_rules} can be applied directly at the level of the Lagrangian, thus giving an off-shell, non-perturbative definition of the double copy.
%\footnote{Remarkably,  Chern-Simons theory {\it automatically} manifests off-shell, Lagrangian-level color-kinematics duality \cite{HenrikCS}.}

\smallskip

\noindent {\it BAS Theory.} The Lagrangian for BAS theory is 
\eq{
\mathcal{L}_{\rm BAS} &=
\tfrac{1}{2} \partial_\mu \phi_{a \ov a} \partial^\mu \phi^{a \ov a} + \tfrac{1}{6} f_{abc} f_{\ov a \ov b \ov c} \phi^{a \ov a} \phi^{b \ov b} \phi^{c \ov c} ,
}{L_BAS}
while the corresponding equation of motion is
\eq{
\Box \phi^{c \ov c} -\tfrac{1}{2} \fddu{a}{b}{c} \fbarddu{a}{b}{c} \phi^{a \ov a} \phi^{b \ov b} =0 .
}{EOM_BAS}
The tree-level four-point off-shell BAS amplitude is
\eq{
-A_{\rm BAS} = \frac{c_s \ov c_s}{s} +\frac{c_t \ov c_t}{t}+\frac{c_u \ov c_u}{u},
}{A_BAS}
where $s=(p_1+p_2)^2$, $t=(p_2+p_3)^2$, $u=(p_3+p_1)^2$, and the color factors are $c_s = \fddu{a_1}{a_2}{b} f_{b a_3 a_4}$, 
$c_t = \fddu{a_2}{a_3}{b}f_{b a_1 a_4}$,
$c_u = \fddu{a_3}{a_1}{b}f_{b a_2 a_4}$,
and likewise for barred color.  

Massless {\it on-shell} kinematics in two dimensions is famously plagued by infrared singularities since all asymptotic states are either left- or right-movers.    For example, for the case of four-point scattering with color-ordered external states, the external momenta exhibit kinematic configurations which we classify as ``split'', where $p_1+p_2 = p_3+p_4=0$ or $p_2+p_3 = p_1+p_4=0$, versus ``alternating'', where $p_3 +p_1 = p_2 + p_4$.     Since either  $s$, $t$, or $u$ is always zero, there is a vanishing Gram determinant, $stu=0$, and propagator exchanges generically exhibit collinear singularities.   

The precise method of infrared regulation---be it going off-shell, introducing a physical mass term to the theory, or analytically continuing away from two dimensions---can yield different answers for nominally classical equivalent theories, and special care must be taken \cite{Tseytlin}.  Nevertheless, the claim of the present paper is that {\it assuming} a particular infrared regulator, our construction can be applied to map any given infrared-regulated theory to a corresponding infrared-regulated double copy theory.

\smallskip

\noindent {\it ZM Theory.} Applying the replacement rules in \Eq{replacement_rules} to the {\it Lagrangian} of BAS theory in \Eq{L_BAS}, we obtain the {\it action} of ZM theory, whose Lagrangian is \cite{ZM, Polyakov, Nappi, Tseytlin}
\eq{
\mathcal{L}_{\rm ZM} &= 
\tfrac{1}{2} \partial_\mu \phi_{a } \partial^\mu \phi^{a } + \tfrac{1}{6} f_{abc}   \phi^{a} \partial_\mu \phi^{b } \tpartial^\mu \phi^{c } .
}{L_ZM}
The resulting equation of motion is
\eq{
\Box \phi^c -\tfrac{1}{2} \fddu{a}{b}{c} \partial_\mu \phi^a \tpartial^\mu \phi^b=0,
}{EOM_ZM}
which can alternatively be obtained from \Eq{EOM_BAS} via \Eq{replacement_rules}. Note that \Eq{EOM_ZM} also encodes the dynamics of SDYM \cite{SDYM1, SDYM2, Chacon:2020fmr}.   

As is well-known \cite{ZM, Nappi, Tseytlin}, ZM theory is classically equivalent to the principal chiral model (PCM), otherwise known as the non-linear sigma model (NLSM) in two dimensions.  In general dimensions, the NLSM is classically defined by
\eq{
\partial_{[\mu} j_{\nu]}^c + \fddu{a}{b}{c}j_\mu^a j_\nu^b =0 \quad {\rm and} \quad \partial^\mu j_\mu^a=0,
}{}
where the former is a pure gauge condition implying that $j_\mu^a\b T_a \sim \b g ^{-1} \partial_\mu \b g$ and the latter is the NLSM equation of motion.  By defining $j_\mu^a =  \epsilon_{\mu\nu} \partial^\nu \phi^a$, we trivially enforce the latter, while the former is equivalent to \Eq{EOM_ZM}.

The three-point Feynman vertex defined by \Eq{L_ZM} is 
\eq{
\includegraphics[trim={0 0 0 0},clip,valign=c,scale=0.7]{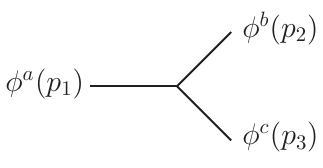} \quad = \quad -if_{abc} \langle 12\rangle,
}{diagram_ZM}
which is fully antisymmetric because off-shell two-dimensional kinematics implies that 
$\langle 12\rangle=\langle 23\rangle=\langle 31\rangle$. 

The tree-level four-point off-shell ZM amplitude is
\eq{
-A_{\rm ZM} = \frac{c_s n_s}{s} +\frac{c_t  n_t}{t}+\frac{c_u n_u}{u},
}{A_ZM}
where the kinematic numerators,
\eq{
n_s =  \langle 12\rangle\langle34\rangle , \quad
n_t =  \langle 23\rangle\langle14\rangle, \quad
n_u =  \langle 31\rangle\langle24\rangle ,
}{num_ZM}
satisfy the off-shell  kinematic Jacobi identity,
$n_s +n_t +n_u=0$, on account of the Schouten identity. Applying the standard color decomposition \cite{DelDuca:1999rs}, the color-ordered ZM amplitude is $A_{\rm ZM}[1234] = \frac{n_s}{s} - \frac{n_t}{t}$.  

For the alternating configuration described previously, $u=s+t = 0$, which implies that $A_{\rm ZM}[1234]$ is free of collinear singularities.  In this case $n_s = -n_t = \langle 12\rangle^2$, so $A_{\rm ZM}[1234]=0$, in accordance with the phenomenon of no-particle production described in \cite{Gabai:2018tmm}.  For the split configurations, $A_{\rm ZM}[1234]$ is non-zero but must be evaluated with some choice of infrared regulator \cite{Tseytlin}.

At loop level, integrands at arbitrary order are mechanically calculated using the Feynman vertex in \Eq{diagram_ZM}.  By construction, all loop-level kinematic Jacobi identities are automatically satisfied, even off-shell.  While enforcing ``global color-kinematics constraints'' is a well-known difficulty in gauge theory starting at two-loops \cite{Bern:2015ooa}, we learn here that there is no obstacle to this for ZM theory at all orders in perturbation theory.

\smallskip

\noindent {\it SG Theory.} \Eq{replacement_rules} maps the ZM Lagrangian in \Eq{L_ZM} to the action of the SG theory, whose Lagrangian is
\eq{
\mathcal{L}_{\rm SG} &=  \tfrac{1}{2} \partial_\mu \phi \partial^\mu \phi + \tfrac{1}{6}   \phi \partial_{\mu}\partial_\nu \phi \tpartial^\mu \tpartial^\nu \phi ,
%\\
%&= \int  \frac{1}{2} \partial_\mu \phi \partial^\mu \phi - \frac{1}{6}   \phi  (\Box \phi \Box \phi - \partial_{\mu\nu} \phi \partial^{\mu\nu} \phi ). 
}{}
and whose equation of motion is
\eq{
\Box \phi -\tfrac{1}{2} \partial_\mu  \partial_\nu \phi \tpartial^\mu \tpartial^\nu \phi=0.
}{EOM_SG}
The three-point Feynman vertex is then
\eq{
\includegraphics[trim={0 0 0 0},clip,valign=c,scale=0.7]{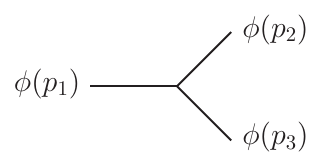} \quad = \quad i \langle 12\rangle^2,
}{}
which is fully permutation invariant.  
Applying either an off-shell or mass regulator for infrared singularities, the on-shell amplitude is
\eq{
-A_{\rm SG} = \frac{n_s^2}{s} +\frac{n_t^2}{t}+\frac{n_u^2}{u} \sim stu %-\tfrac{9}{16} stu=
=0,
}{A_SG}
which is proportional to the Gram determinant and thus vanishes in two dimensions.  This  reflects the fact that the SG is field-redefinition equivalent to a two-dimensional free theory \cite{sGalDuality,GenGalDuality}.

\Section{Masses and Higher-Dimension Operators.}  Thus far we have only considered those theories which have historically appeared in the amplitudes-level double copy \cite{BCJReview}. Our construction extends far more broadly, however.  In particular, the color-kinematic replacement rules in \Eq{replacement_rules} can be applied to {\it any operator} that does not have {\it i}) a closed loop of color structure constants, nor {\it ii}) multiple color traces.  In both cases, the third line of \Eq{replacement_rules} will induce ill-defined or non-local integrals over the volume of spacetime which enter into the Lagrangian.
The mildness of the restrictions {\it i}) and {\it ii}) means that a very large class of operators manifestly obey color-kinematics duality, in sharp contrast to the typical intuition that almost all operators will fail the duality.

By these rules, mass terms are perfectly fine and trivially double copy via the same color-kinematic replacements as the kinetic terms.  These mass terms serve only to change the propagator denominators.

\Eq{replacement_rules} can also be implemented for an infinite class of higher-dimension operators.
For example, consider the higher-dimension operator in BAS theory,
${\cal O}_{\rm BAS}  = f_{ a b e} f_{ c d  e} f_{\ov a\ov b\ov e} f_{\ov c\ov d \ov e}  \phi^{a\ov a} \phi^{b\ov b} \phi^{c\ov c}\phi^{d\ov d}$,
where both the color and dual color structures are single-trace.  Applying \Eq{replacement_rules} to dual color, we obtain the spacetime integral of ${\cal O}_{\rm ZM} =  f_{ a b e} f_{ c d  e}   \partial_\mu \phi^a \tpartial^\mu \phi^b \partial_\nu \phi^c \tpartial^\nu \phi^d$, which is the color-kinematic dual  operator in ZM theory. 
Then applying \Eq{replacement_rules} to the remaining color structures, we obtain the spacetime integral of
%\eq{
${\cal O}_{\rm SG}  = \partial_\mu \partial_\nu \phi \tpartial^\mu \tpartial^\nu \phi \partial_\rho\partial_\sigma \phi \tpartial^\rho \tpartial^\sigma \phi$,
%-\tfrac{1}{2}   \partial_\rho \partial_\mu \phi \tpartial^\rho \partial_\nu \phi\partial_\sigma \partial^\mu \phi  \tpartial^\sigma\partial^\nu \phi \\
%&=-\tfrac{1}{2} \left(  [\Pi^4]- [\Pi^2]^2 \right) \quad {\rm where} \quad \Pi_{\mu\nu} = \partial_\mu \partial_\nu \phi.
%}{}
which is the color-kinematic dual operator in the SG theory.

Now consider
%\eq{
${\cal O}_{\rm BAS}'  = g_{ac} g_{bd}f_{\ov a\ov b\ov e} f_{\ov c\ov d \ov e}  \phi^{a\ov a} \phi^{b\ov b} \phi^{c\ov c}\phi^{d\ov d}$,
%}{}
which is double-trace in color and single-trace in dual color.  Applying \Eq{replacement_rules} to the latter, we obtain
%\eq{
${\cal O}_{\rm ZM}'  =    \partial_\mu \phi_a \tpartial^\mu \phi_b \partial_\nu \phi^a \tpartial^\nu \phi^b$.
%\\
%& = (\partial_\mu \phi^a \partial^\mu \phi^a)^2-(\partial_\mu \phi^a \partial^\mu \phi^b)^2,
%}{}
Since the resulting operator is double-trace in color, it cannot be further double copied via \Eq{replacement_rules} without generating an additional integral over all of spacetime.

\Section{Fundamental BCJ Relation.} Our Lagrangian-level formulation of the double copy does not preserve the fundamental Bern-Carrasco-Johansson (BCJ) relation \cite{BCJ1, BCJ2} nor the so-called minimal rank condition \cite{ElvangPhi3}.  Ultimately, this is not so surprising because the fundamental BCJ relation is literally equivalent to the conservation equation for the kinematic current in theories with purely cubic interactions \cite{CCK, AdS_CK}, and our generalized double copy construction allows for quartic and higher interactions.  

Crucially, failure of the minimal rank condition implies that our framework should be interpreted as a generalization of the color-kinematic dual formulation of the double copy \cite{BCJ1, BCJ2}, which is built upon the kinematic Jacobi identities, rather than the Kawai-Lewellen-Tye (KLT) formulation \cite{KLT}, which relies on relations amongst color-ordered amplitudes.

As an example, consider BAS theory deformed by a mass and the higher-dimension operator defined earlier,
\eq{
\mathcal{L} = \mathcal{L}_{\rm BAS} -\tfrac{m^2}{2}  \phi_{a \ov a} \phi^{a \ov a} + \tfrac{\tau}{16} {\cal O}_{\rm BAS}.
}{}
For the moment, let us work in general dimensions, where infrared divergences are absent.  The matrix of doubly color-ordered amplitudes is
\eq{
&H(m,\tau) = \left(
\begin{array}{cc}
A[1234 \vert 1234] & A[1234 \vert 1324]\\
A[1324 \vert 1234] & A[1324 \vert 1324] 
\end{array}
\right) \\
&= -\left(
\begin{array}{cc}
\frac{1}{s-m^2} +\frac{1}{t-m^2}-{\scriptstyle\tau} &-\frac{1}{t-m^2}+\tfrac{\tau}{2}\\
-\frac{1}{t-m^2}+\tfrac{\tau}{2}& \frac{1}{t-m^2} +\frac{1}{u-m^2}-{\scriptstyle \tau}
\end{array}
\right).
}{H_m_tau}
The minimal rank condition holds
for pure BAS theory in general dimensions since $\det H(0,0) = 0$ on-shell.  Howver, it fails in the presence of masses \cite{Johnson:2020pny, Momeni:2020vvr} and higher-dimension operators \cite{ElvangPhi3},
\eq{
\det H(m,0) &= \tfrac{m^2}{(s-m^2)(t-m^2)(u-m^2)} \\
\det H(0,\tau) &= -{\scriptstyle \tau} \left(\tfrac{1}{s} +\tfrac{1}{t}+\tfrac{1}{u}\right) + \tfrac{3\tau^2}{4}.
}{detH}
Evaluating these expressions for  two-dimensional kinematics, we encounter the usual annoyances of infrared divergences, but irrespective of choice of regulator, the above determinants are still non-zero.

\Section{Integrable Models.} Since ZM theory is classically equivalent to the PCM, it is similarly integrable \cite{ZM, Polyakov, Curtright}. Moreover, ZM theory maps to the SG under the double copy, so we will see that the latter is also integrable.
Note that mapping the integrability of one theory to another would not be possible with the standard amplitudes-level double copy because the integrability conditions are expressed in terms of currents and off-shell fields.

As a brief review, integrability is achieved by casting the equations of motion into the form of the {\it Lax equation}, $\dot{L} = [M,L]$, where the operators $L$ and $M$ constitute a {\it Lax pair} \cite{Lax:1968fm,Beisert, Torrielli:2016ufi,IntegrabilityBook1,IntegrabilityBook2}.
By virtue of this form of the equations of motion, the eigenvalues of $L$ are conserved charges.
A familiar Lax pair is the Hamiltonian together with an observable in the Heisenberg picture.
In two dimensions, integrability requires infinitely many charges where the infinitude of Lax pairs is parameterized by a {\it spectral parameter} $\lambda$.
The Lax pair comes from a Wilson line and a gauge field, the {\it Lax connection}, where flatness of the gauge connection yields the Lax equation.
%More concretely, the charges will be built from a Wilson line which is in turn built from a flat gauge potential called a {\it Lax connection}.
%More concretely, the conserved charges will come from the series coefficients in $\lambda$ of a Wilson line.
%Path independence of the Wilson line requires it to be built from a flat gauge field, a {\it Lax connection}, whose flatness is guaranteed by the original equation of motion.
%More details on integrability can be found in \Reff{Beisert, Torrielli:2016ufi}.

%\noindent {\it Color Currents.} The Lagrangian of BAS theory is invariant under the color and dual color symmetry transformations
%whose corresponding conserved currents are
%\eq{
%{\cal J}_\rho^{c} = g_{\ov a \ov b} \fddu{a}{b}{c} \phi^{a\ov a} \overset{\leftrightarrow}{\partial_\rho} \phi^{b\ov b}\quad {\rm and} \quad
%{\cal J}_\rho^{\ov c} = g_{ab} \fbarddu{a}{b}{c} \phi^{a\ov a} \overset{\leftrightarrow}{\partial_\rho} \phi^{b\ov b}.
%}{}
%Under the color-kinematic replacement in \Eq{replacement_rules}, the latter maps to the kinematic current of ZM theory,
%\eq{
%{\cal J}_\rho^{\ov c} \quad \rightarrow \quad {\cal J}_\rho =\partial_\mu \phi_{a} \overset{\leftrightarrow}{\partial_\rho} \tpartial^\mu  \phi^{a},
%}{}
%which is conserved via the equations of motion,
%\eq{
%\partial^\rho {\cal J}_\rho = \partial_\mu \phi_{a} \overset{\leftrightarrow}{\Box} \tpartial^\mu  \phi^{a} =0,
%}{}
%and enforces the kinematic Jacobi identity and is closely related to the energy-momentum tensor \cite{CCK}.  

\smallskip

\noindent {\it Integrability of ZM Theory.} 
Let us review the integrability properties of ZM theory \cite{ZM, Polyakov, Curtright}.  To begin, we define the Lax connection \cite{ZM, Curtright, PCMReview1, PCMReview2},
\eq{
\b A_\mu = \tfrac{1}{1-\lambda^2} (\tpartial_\mu \b\phi +\lambda \partial_\mu \b \phi),
}{A_Lax}
whose corresponding field strength,
\eq{
\b F_{\mu\nu} = \partial_{[\mu} \b A_{\nu]}+ [\b A_{\mu},\b A_{\nu}] =0,
}{F_Lax}
vanishes for all values of the parameter $\lambda$ due to the ZM equation of motion in \Eq{EOM_ZM}.  Since the Lax connection is pure gauge, we can construct the Wilson line,
\eq{
&\b W(x) = P \exp\left[ -\int^x dx'^\mu  \b A_\mu(x') \right] \\
&= 1-\int^x dx' \b A(x')+ \int^{x} dx' \b A(x') \int^{x'} dx'' \b A(x'')+\cdots,
}{}
which is path-independent and satisfies
$D_\mu \b W = \partial_\mu \b W +   \b A_\mu \b W=0$.
Next, we define the Lax current  \cite{ZM, Polyakov, Curtright}
\eq{
\b J_\mu = \tpartial_\mu \b W = \sum_{k=0}^\infty \lambda^{-k} \b J_\mu^{(k)},
}{J_tower}
which furnishes an infinite tower of currents, including
\eq{
 \b J_\mu^{(1)} &= \tpartial_\mu \b \phi ,\quad 
 %\b J_\mu^{(2)} = \partial_\mu \b \phi -\tfrac{1}{2} [\b \phi, \tpartial_\mu \b \phi]+ \tfrac{1}{2} \tpartial_\mu (\b %\phi \, \b \phi) \\
  \b J_\mu^{(2)} = \partial_\mu \b \phi +  \tpartial_\mu \b \phi \, \b \phi  \\
   \b J_\mu^{(3)} &= \tpartial_\mu \b \phi +\partial_\mu \b \phi \, \b \phi + \tpartial_\mu \b \phi \int^x dx' \tpartial \b \phi + \tpartial_\mu \b \phi \int^x dx' \partial \b \phi \, \b \phi ,
}{currents_ZM}
which become increasingly non-local at higher orders.
On the support of the equations of motion in \Eq{EOM_ZM}, these currents are conserved, so $\partial^\mu  \b J_\mu =\partial^\mu  \b J_\mu^{(k)} = 0$.

\smallskip

\noindent {\it Integrability of SG Theory.} Applying the color-kinematics replacement in \Eq{replacement_rules} to \Eq{A_Lax} and \Eq{F_Lax} we obtain the Lax connection for SG theory,
\eq{
A_\mu = \tfrac{1}{1-\lambda^2} (\tpartial_\mu \phi +\lambda \partial_\mu  \phi),
}{}
whose corresponding field strength is also vanishing,
\eq{
F_{\mu\nu} = \partial_{[\mu}  A_{\nu]}+  \partial_\rho A_{\mu}  \tpartial^\rho A_{\nu} =0.
}{F_Lax_SG}
Meanwhile, the Wilson line maps from a color matrix to a diffeomorphism via
\eq{
& \b W(x) = P \exp \left[ \int^x dx'^\mu \tpartial^\nu A_\mu \partial_\nu \right] \\
&= 1+ \int^x dx'^\mu \tpartial^\nu A_\mu \partial_\nu+\int^x dx'^\mu   \tpartial^\nu A_\mu \tpartial^\rho A_\nu \partial_\rho  +\cdots \\
&= \int^x dx'^\mu K_\mu^\nu \partial_\nu \quad {\rm where} \quad  K_\mu^\nu =  (\delta_\mu^\nu - \tpartial^\nu A_\mu)^{-1}.
}{}
As per \Eq{J_tower}, the Lax current for the SG theory is
\eq{
\b J_\mu = \epsilon_{\mu\nu}  \tpartial^\rho A^\nu  K_\rho^\sigma \partial_\sigma,
}{J_tower_SG}
which is conserved since
\eq{
\partial^\mu \b J_\mu &=-\tilde \partial^\nu (  \tpartial^\rho A_\nu K^{\sigma}_\rho )\partial_\sigma =-\tilde \partial^\nu (  K^{\rho}_\nu \tpartial^\sigma A_\rho  )\partial_\sigma =0,
}{}
where $ K^{\rho}_\nu \partial_\mu A_\rho = \partial_\nu A_\mu$ follows directly from \Eq{F_Lax_SG}. The series expansion of \Eq{J_tower_SG} yields an infinite tower of conserved currents in the SG theory which include
\eq{
 \b J_\mu^{(1)} &= \tpartial_\mu \partial_\nu \phi \tpartial^\nu ,\quad 
  \b J_\mu^{(2)} = \partial_\mu \partial_\nu \phi \tpartial^\nu +  \tpartial_\mu \partial_\nu  \phi  \, \tpartial^\nu \partial_\rho  \phi \tpartial^\rho ,
}{currents_SG}
and can also be obtained trivially from the currents of ZM theory in \Eq{currents_ZM} by applying the color-kinematic replacement rules in \Eq{replacement_rules}.

\Section{Non-perturbative Solutions.} \Eq{algebra_equivalence} implies a {\it non-perturbative} map between the classical solutions of the equations of motion of BAS, ZM, and SG theory.

Since the SG theory is field-redefinition equivalent to a two-dimensional free theory \cite{sGalDuality,GenGalDuality,NonPertSGDual,JaraSGDual}, any arbitrary configuration of left- and right-moving wave-packets will pass through each other unscathed even though the collision itself will be highly non-linear and non-perturbative.  Thus if we restrict to scattering on a spatial circle of circumference $2\pi$, then the time evolution will be similarly $2\pi$ periodic.   Since every classical solution of the SG theory effectively resides on a spacetime torus, it can be expressed as a double discrete Fourier transform,
\eq{
\phi(x) = \sum_{p\in \mathbb{Z}\times \mathbb{Z}} e^{ ipx} \tilde \phi(p)=  \sum_{p\in \mathbb{Z}_N\times \mathbb{Z}_N} e^{ ipx} \tilde \phi(p) + {\cal O}(\tfrac{1}{N}),
}{phi_SG}
where the corrections to the right-hand side are negligible as long as the field does not vary on distances shorter than $\tfrac{1}{N}$, which is always true for sufficiently large $N$.

\begin{figure*}
\begin{center}
  \includegraphics[width=1.9 \columnwidth]{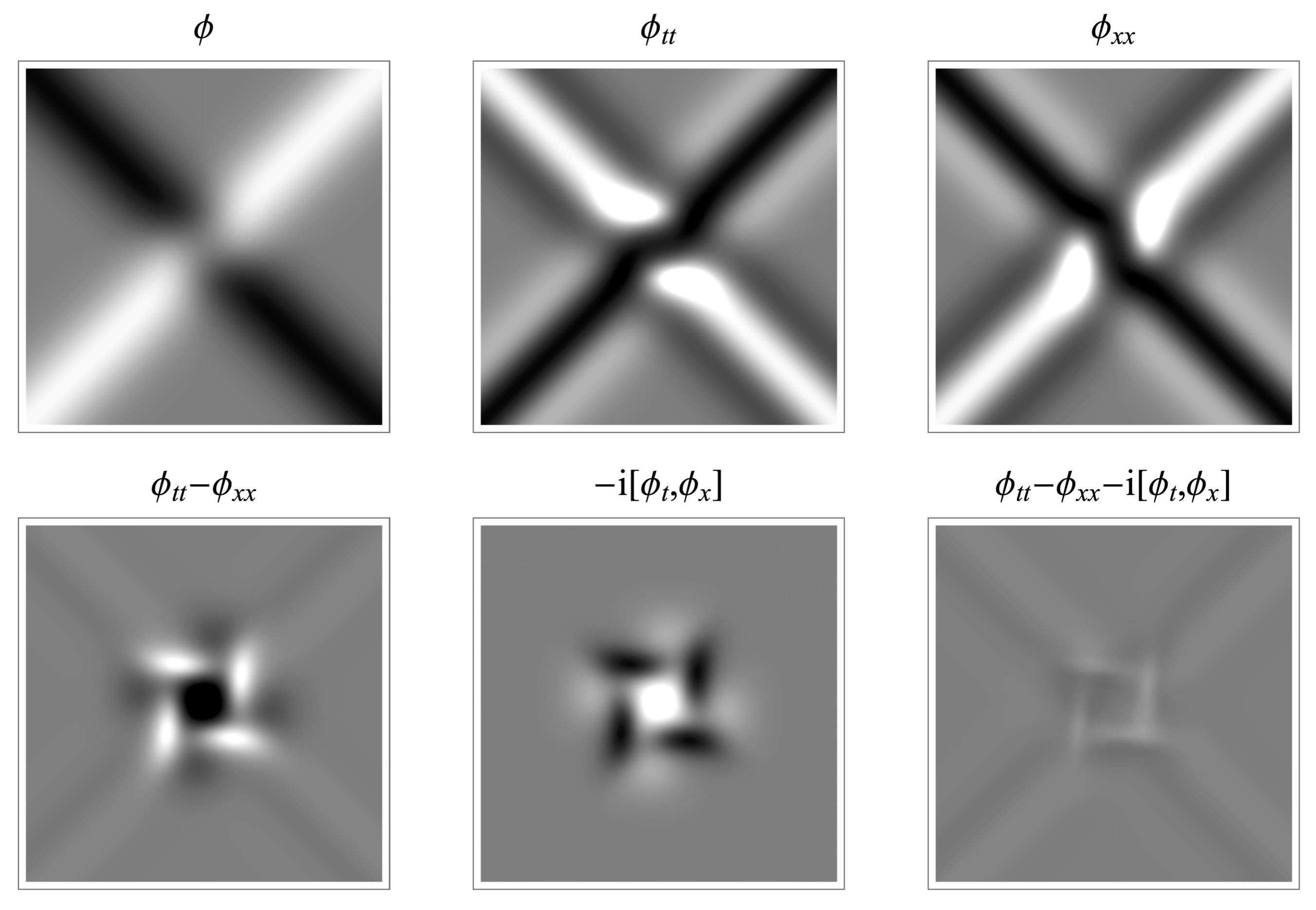}
  \end{center}
\caption{We numerically solve the SG equations of motion in \Eq{EOM_SG} for a pair of colliding Gaussian wave-packets.  The discrete Fourier transform of this solution, defined in \Eq{phi_SG}, is inserted into \Eq{phi_ZM} to obtain a putative matrix-valued solution of ZM theory. The above density plots characterize this ZM configuration, where
the horizontal/vertical axes denote space/time and lighter/darker colors denote positive/negative field values. Each panel depicts a different matrix-valued, spacetime-dependent operator, ${\cal O} = \phi, \phi_{tt}, \phi_{xx}$, etc., where the subscripts denote derivatives.  Each operator is visualized by plotting its projection onto a single component, ${\rm tr}({\cal O} \b T_0)$, where $\b T_0 = \sum_p \b T_p$.  Each term in the ZM equations of motion in \Eq{EOM_ZM} is non-zero, but they nevertheless cancel to high precision in the final panel.  
%Since a pure left- or right-mover is an exact solution of the SG theory, the ZM field configuration shown above deviates from that of a free theory only near the collision.  
These results were obtained for $U(N)$ with $N=499$. See \url{https://bit.ly/3OdGIo4} for an animation of this scattering process.
 }
\label{scattering}
\end{figure*}

We now construct a dual field configuration in ZM theory defined for $U(N)$,
\eq{
\phi^a(x) \, \b T_a = \sum_{p\in \mathbb{Z}_N\times \mathbb{Z}_N} e^{ i px} \tilde \phi(p) \, \b T_p,
}{phi_ZM}
which is literally the SG solution under the replacement $e^{ipx} \rightarrow e^{ipx} T_p$.  It is straightforward to see that \Eq{phi_ZM} {\it automatically satisfies} the ZM equations of motion in \Eq{EOM_ZM} up to $\tfrac{1}{N}$ corrections, since the commutator in \Eq{Tp} and \Eq{fppp_color} yields a  color structure constant that exactly transforms the interaction vertex of ZM into that of SG theory.  Repeating this procedure, we obtain
\eq{
\phi^{a \ov a} (x) \, \b T_a \otimes \b T_{\ov a} = \sum_{p\in \mathbb{Z}_N\times \mathbb{Z}_N} e^{ i px} \tilde \phi(p) \, \b T_p \otimes \b T_p,
}{phi_BAS}
which is a classical solution of BAS theory.

Remarkably, the above analytic construction can be verified {\it numerically}, as described in \Fig{scattering}. Using the double copy replacement, we map a numerical solution of SG theory onto a corresponding matrix-valued field configuration of ZM theory, which is then shown to satisfy the ZM equations of motion to high precision.

Note that ${\it every}$ solution of the SG theory maps to a dual solution in ZM theory but the converse is not true.  This is not actually surprising given what is known from scattering: {\it every} gravity amplitude maps to a gauge theory amplitude with very specific color structures which are precisely chosen to be certain kinematic numerators.  On the other hand, a generic gauge theory amplitude with arbitrary color structures will not have any interpretation as a gravity amplitude.

That the SG is secretly free certainly detracts from the miracle of  a non-perturbative mapping in this context.  However, recall that very general deformations of BAS and ZM theory---for example including masses or higher-dimension operators---also double copy mechanically into analogous deformations of the SG theory.  Non-perturbative solutions of this much larger class of non-free theories will also exhibit the non-perturbative double copy defined in \Eq{phi_SG}, \Eq{phi_ZM}, and \Eq{phi_BAS}.

\Section{Generalization using the Moyal Algebra.} We observed in \Eq{algebra_equivalence} that the $N\rightarrow \infty$ limit of $U(N)$ yields the diffeomorphism algebra. What about finite $N$?
In this case the continuum version of \Eq{fppp_color} is the Moyal algebra \cite{Moyal:1949sk}, 
\eq{
\fddu{p_1}{p_2}{p_3} = -\tfrac{1}{\alpha'} \sin( \alpha' \langle 12\rangle) \delta^2(p_i +p_j-p_k),
}{fppp_ZM_stringy}
which is the unique deformation of the Poisson algebra \cite{Fletcher:1990ib} encoding an infinite tower of higher-dimension corrections to the original kinematic structure constant in \Eq{fppp_ZM}.
Here we have defined a new coupling constant $\alpha' \sim \frac{2\pi}{N}$.
%We refer to \Eq{fppp_ZM_stringy} as the stringy kinematic structure constant, not because of any explicit connection to string theory, but rather since it encodes an infinite tower of higher-dimension corrections to the original kinematic structure constant in \Eq{fppp_ZM}.
At the level of fields, the generalized color-kinematic replacement rule is
\eq{
 \fddu{a}{b}{c} V^a W^b &\quad \rightarrow \quad  \tfrac{1}{\alpha'} \sin\left(\alpha' \partial_{V} \tpartial_W \right)  V W ,
}{replacement_rules_stringy}
where the subscripts denote which fields the derivatives act upon.  Under this substitution, BAS theory maps to
\eq{
\mathcal{L}_{\textrm{ZM}, \alpha'} &= 
\tfrac{1}{2} \partial_\mu \phi^{a } \partial^\mu \phi^{a } + \tfrac{1}{6\alpha'} f_{abc}   \phi^{a} \sin\left(\alpha' \partial_{\phi^b} \tpartial_{\phi^c} \right) \phi^{b }  \phi^{c } ,
}{L_ZMStringy}
a Moyal-deformed variation of ZM theory which has also appeared in the context of SDYM \cite{Chacon:2020fmr}.

The corresponding three-point Feynman vertex is
\eq{
\includegraphics[trim={0 0 0 0},clip,valign=c,scale=0.7]{ZMVert} \quad = \quad -\tfrac{i}{\alpha'}f_{abc} \sin(\alpha'  \langle 12\rangle),
}{V3_ZM_stringy}
which is totally antisymmetric because of two-dimensional kinematics.
The resulting four-point amplitude is given by \Eq{A_ZM} with the numerators
\eq{
n_s &=  \tfrac{1}{\alpha'^2} \sin(\alpha' \langle 12)\rangle  \sin(\alpha' \langle34\rangle) \\
n_t &=   \tfrac{1}{\alpha'^2}\sin(\alpha'  \langle 23\rangle)  \sin(\alpha' \langle14\rangle) \\
n_u &=  \tfrac{1}{\alpha'^2} \sin(\alpha' \langle 31\rangle)  \sin(\alpha' \langle24\rangle) .
}{num_ZM_stringy}
Remarkably, these satisfy the kinematic Jacobi identity for any value of $\alpha'$, so for example
\eq{
\sin  \langle 12\rangle\sin  \langle 34\rangle+\sin  \langle 23\rangle\sin  \langle 14\rangle+\sin  \langle 31\rangle\sin  \langle 24\rangle =0,
}{}
for arbitrary off-shell two-dimensional kinematics.

The generalized replacement rule in \Eq{replacement_rules_stringy} can be reapplied to ZM to generate a deformation of SG theory that includes a fixed tower of higher-dimension corrections, analogous to the infinite tower of corrections to self-dual gravity in \Reff{Chacon:2020fmr}.

\Section{Future Directions.} 
% The double copy is an extremely potent computational tool for tackling the dramatic increase in complexity from gauge theory to gravity. 
% %The double copy is an extremely potent tool mapping gauge theory to gravity. The mechanical power of  the duality stems from the dramatic increase in complexity from gauge theory to gravity.
% While more and more applications are found for the double copy, it remains fundamentally unclear {\it why} it works. The duality between these theories points to hidden structure that is invisible at the level of the off-shell actions.  
% %This is the type of question one might try to explore from an off-shell Lagrangian level perspective, and if answered would have bearing on all double copy constructions.
% In this letter we have presented the first Lagrangian level double copy between interacting theories which provides insight into the underlying structure behind color-kinematics duality.

The double copy is an extremely potent computational tool but it is fundamentally unclear {\it why} it works.
Our results mark a radical departure from the status quo of the double copy in several ways.
Typical theories that admit color-kinematics duality have a single coupling constant, massless particles, square in any spacetime dimension, only double copy on-shell, and all of this is only provable at tree-level \cite{BCJReview}.
On the other hand we have presented an enormous class of scalar theories with arbitrary Wilson coefficients and masses that square off-shell (to all orders in perturbation theory) in two dimensions. A Lagrangian formulation coupled with an understanding of the algebra mapping also broadens the scope of the double copy to include Wilson lines, currents, and non-perturbative (non-abelian) classical solutions.

The present work leaves several avenues for further inquiry.  %First and foremost is the question of generalization to higher dimensions.
In general dimensions, the kinematic algebra for the NLSM is that of volume-preserving diffeomorphisms \cite{CCK}.
Generalizing this tree-level observation to the full loop-level action is an open problem.
The two dimensional results presented here suggest that this generalization may be possible, at least in principle.
While we have found an enormous class of operators that double copy, it may be possible to enlarge the space even further by overcoming the restrictions {\it i}) and {\it ii}) given above.

Finally, it would also be interesting to apply our approach to gauge theory and gravity in two dimensions and beyond.
The kinematic algebra for gauge theory \cite{CCK}, even at tree level, is not as well understood as for the NLSM.
However, the self-dual sector of Yang-Mills theory has a simple kinematic algebra so it may be possible to systematically perturb away from the self-dual sector \cite{SDYM1, SDYM2}.

\medskip
\Section{Acknowledgments.}
We are grateful to J.J.~Carrasco, Henrik Johansson, Ricardo Monteiro, and Donal O'Connell for insightful comments on the draft.
C.C., J.P.-M., and N.S. are supported by the DOE under grant no.\ DE-SC0011632 and by the Walter Burke Institute for Theoretical Physics. 
J.M. is supported in part by the DOE under contract DE-SC0021485 as well as the Northwestern University Amplitudes and Insight group.
Numerical simulations were performed on the Hoffman2 Cluster at the Institute for Digital Research and Education at UCLA. J.P.-M.~is also supported in part by the NSF under Grant No.~NSF PHY-1748958 and would like to thank the Mani L. Bhaumik Institute for Theoretical Physics and the Kavli Institute for Theoretical Physics for hospitality.

\bibliographystyle{utphys-modified}
\bibliography{flatbib}

\end{document}